\documentclass[10pt]{article}

\usepackage{amsmath,amsfonts,amssymb}
\usepackage{graphicx}
\usepackage{authblk}

\title{Photo-induced second-order nonlinearity in stoichiometric silicon nitride waveguides.}

\author[1]{Marco A.G. Porcel}
\author[1]{Jesse Mak}
\author[1]{Caterina Taballione}
\author[1]{Victoria K. Schermerhorn}
\author[1,2]{J\"{o}rn P. Epping}
\author[1]{Peter J.M. van der Slot \thanks{Corresponding author: \textit{p.j.m.vanderslot@utwente.nl}}}
\author[1]{Klaus-J. Boller}
\affil[1]{Laser Physics and Nonlinear Optics, MESA+ Institude for Nanotechnology, University Twente, The Netherlands}
\affil[2]{Currently with LioniX International BV., Enschede, The Netherlands}

\begin{document}
 
\maketitle
 
\begin{abstract}
We report the observation of second-harmonic generation (SHG) in stoichiometric silicon nitride waveguides grown via low-pressure chemical vapor deposition (LPCVD). Quasi-rectangular waveguides with a large cross section were used, with a height of 1~$\mu$m and various different widths, from 0.6 to 1.2~$\mu$m, and with various lengths from 22 to 74~mm. Using a mode-locked laser delivering 6-ps pulses at 1064~nm wavelength with a repetition rate of 20~MHz, 15\% of the incoming power was coupled through the waveguide, making maximum average powers of up to 15~mW available in the waveguide depending on the waveguide cross section. Second-harmonic output was observed with a delay of minutes to several hours after the initial turn-on of pump radiation, showing a fast growth rate between 10$^{-4}$ to 10$^{-2}$ s$^{-1}$, with the shortest delay and highest growth rate at the highest input power. After this first, initial build-up (observed delay and growth), the second-harmonic became generated instantly with each new turn-on of the pump laser power. Phase matching was found to be present independent of the used waveguide width, although the latter changes the fundamental and second-harmonic phase velocities. We address the presence of a second-order nonlinearity and phase matching, involving an initial, power-dependent build-up, to the coherent photogalvanic effect. The effect, via the third-order nonlinearity and multiphoton absorption leads to a spatially patterned charge separation, which generates a spatially periodic, semi-permanent, DC-field-induced second-order susceptibility with a period that is appropriate for quasi-phase matching. The maximum measured second-harmonic conversion efficiency amounts to 0.4\% in a waveguide with $0.9\times1$~$\mu$m$^2$ cross section and 36~mm length, corresponding to 53~$\mu$W at 532~nm with 13~mW of IR input coupled into the waveguide. The according $\chi^{(2)}$-susceptibility amounts to 3.7~pm/V, as retrieved from the measured conversion efficiency.
\end{abstract}


\bibliography{bibliography_complete} 
\bibliographystyle{osajnl}

\section{Introduction}

Photonic integrated circuits (PICs) realized with various different material platforms, \textit{e.g.}, SOI (silicon on insulator)~\cite{pavesi_silicon_2004, tekin_optical_2016}, InP~\cite{hunsperger_integrated_2013, smit_introduction_2014}, polymers~\cite{hunsperger_integrated_2013}, SiO$_2$~\cite{hunsperger_integrated_2013, tekin_optical_2016} and SiN~\cite{bauters_ultra-low-loss_2011, marpaung_integrated_2013}, have gained a tremendously growing importance in modern photonic technologies. This is due to a huge variety of emerging applications in high-throughput communications \cite{vlasov_high-throughput_2008}, optical sensing \cite{monat_integrated_2007} and the life sciences, the latter, specifically, when involving visible light \cite{yang_atomic_2007, calafiore_holographic_2014}. Semiconductor waveguide platforms offer the strongest first-order optical interactions, such as required for light generation, amplification or photo detection. Dielectric waveguide platforms, on the other hand, provide lowest propagation loss and tolerate high intensities, thereby enabling various types of optical functionalities via nonlinear optical interactions.

A most prominent example is photonic circuits fabricated from stoichiometric Si$_{3}$N$_{4}$ grown via low-pressure chemical vapor deposition (LPCVD), embedded in a $\text{SiO}_{2}$ cladding. This platform offers a unique combination of adjustable properties, \textit{i.e.}, record-low propagation loss ($<$0.001~dB/cm) \cite{bauters_ultra-low-loss_2011} such as for high-Q resonators \cite{spencer_integrated_2014}, a wide spectral range of optical transparency (from about 310~nm throughout the entire visible spectrum up to 5.5~$\mu$m~\cite{luke_broadband_2015}), and a high index contrast to achieve tight mode confinement via fabrication of relatively thick waveguide cores~\cite{epping_high_2015}. This platform has also reached a considerable degree of maturity, allowing two-dimensional tapering \cite{worhoff_triplex:_2015} for low-loss fiber coupling or realizing hybrid lasers \cite{fan_hybrid_2014, fan_optically_2016}. The platform supports also a wide range of optical functionalities~\cite{spencer_integrated_2014, roeloffzen_silicon_2013} by offering various different core cross sections for waveguide dispersion engineering, and offers a highly reproducible material dispersion \cite{worhoff_triplex:_2015, epping_high_2015}. Regarding third-order nonlinearities, the recent demonstration of the broadest-ever optical spectrum generated on a chip, 495~THz when pumped at 1064~nm \cite{epping_-chip_2015}, shifted more towards the mid-infrared when pumped at 1550~nm, while still maintaining a bandwidth of 453~THz~\cite{porcel_two-octave_2017}. This has opened a wide prospective towards four-wave mixing \cite{epping_integrated_2013}, frequency comb generation \cite{pfeifle_coherent_2014} and all-optical switching \cite{hellwig_ultrafast_2015}.  

Additional functionalities that can be used with lower field strengths would become accessible if this platform would as well provide a second-order nonlinear response, for instance for electro-optic modulation, second-harmonic generation (SHG), or parametric down-conversion. Such nonlinearity, having as its signature a non-zero second-order susceptibility, is also of interest for generating quantum correlated photon pairs directly within reconfigurable time-bin entanglement circuits~\cite{xiong_compact_2015}.

A fundamental precondition for making use of such second-order nonlinearity is, however, that the material provides a non-inversion symmetric structure. With the discussed silicon nitride platform, this is a problem because the involved materials, Si$_{3}$N$_{4}$ and SiO$_2$, are amorphous and thus inversion symmetric. On the other hand, there have been two reports on second-order response in related amorphous SiN-type materials. Specifically, second-harmonic generation has been observed in SiN waveguides (fabricated with plasma enhanced chemical vapor deposition at lower temperatures \cite{ning_strong_2012, levy_harmonic_2011}), and in silicon-enriched SiN films (fabricated with RF sputtering~\cite{kitao_investigation_2014}). Such differences are important to note because both the stoichiometric ratio as well as the type of fabrication process have a strong influence on the size of the optical bandgap, the propagation losses and the third-order nonlinear response~\cite{torres-company_optical_2014}. The stoichiometric, low-loss Si$_{3}$N$_{4}$ material waveguides described above have not been investigated so far for their second-order nonlinear response, thus leaving open whether functionalities based on a non-zero $\chi^{(2)}$ susceptibility can be realized.  

Here we present the first observation of a second-order response in LPCVD-grown, stoichiometric Si$_{3}$N$_{4}$ waveguides, using second-harmonic generation. Employing a mode-locked laser at a wavelength of 1064~nm delivering 6.2~ps pulses and coupling an average power of 13~mW into a waveguide with $0.9\times1$~$\mu$m$^2$ cross-section and 36 mm length, a second-harmonic (SH) output of 53~$\mu$W was reached, which corresponds to a conversion of 0.4\%, with an effective second-order nonlinear susceptibility of 3.7~pm/V. We observed that in order to generate a SH output, an optical initialization process is required, similar to what had been observed earlier in glass fibers~\cite{osterberg_dye_1986,margulis_second-harmonic_1988}. The process involves exposing the waveguides with the in-coupled IR pulses at mW power levels over time intervals between several minutes to hours, depending on the infrared power.

\section{Experimental procedure and results}

In our experimental approach we aimed on demonstrating second-harmonic generation based on modal phase matching ~\cite{suhara_waveguide_2003}. This requires the identification of proper waveguide dimensions that provide the same effective refractive index (waveguide index) for propagation at the fundamental and second-harmonic frequencies. To calculate the effective refractive index dispersion, $n_{\mathrm{eff}}(\omega)$, as function of the width and height of the waveguide core we use a fully vectorial finite-element mode solver~\cite{_comsol_1986} with the Sellmeier dispersion data given by Luke \textit{et al.}~\cite{luke_broadband_2015} for the wavelength range of the available  pump laser (around 1.064~$\mu$m wavelength) and its second-harmonic (around 532~nm). For maximizing the waveguide-internal intensity via a strong confinement, enlarged core cross-sections are considered with a height and width around 1~$\mu$m. The two-dimensional step-index profile used for the calculations is based on the actual waveguide shape available from SEM images ~\cite{epping_high_2015}.

\begin{figure}[htbp]  
	\includegraphics[width=\linewidth]{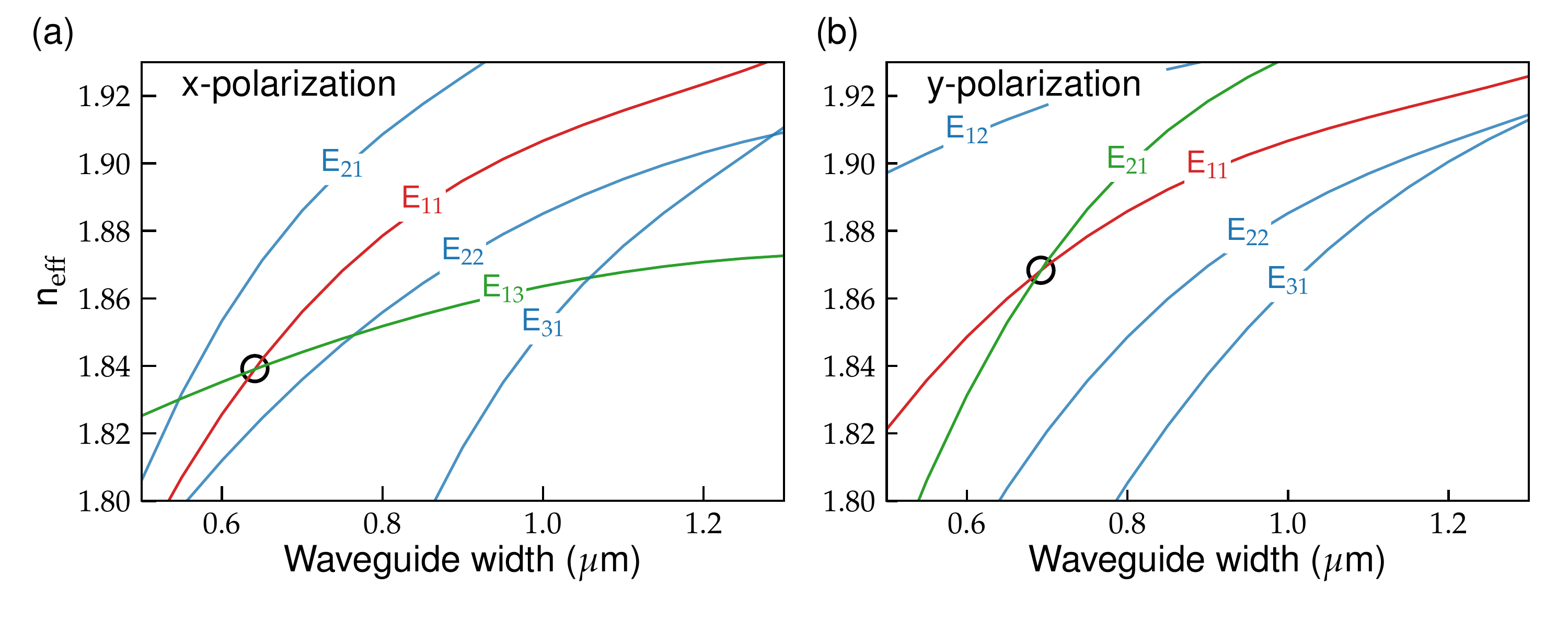} 
	\caption{\label{fig:shg_phasematching} \textbf{(a)} Calculated effective refractive index \textit{vs.} waveguide width, $w$,  for the IR fundamental mode (E$_{11}$) with quasi-horizontal polarization (red curve), and for various transverse modes at 532 with quasi-horizontal polarization. The waveguide height is fixed at 1~$\mu$m. \textbf{(b)} Analogous calculation of dispersion for quasi-vertically polarized modes. The black circles indicate where modal phase matching is expected, namely with $w=0.65$~$\mu$m or 0.68~$\mu$m.}
\end{figure}

Figure~\ref{fig:shg_phasematching} gives an overview of the calculated effective index, $n_{\mathrm{eff}}$, \textit{vs.} the waveguide width, $w$, for a constant waveguide height, $h$, of 1~um for two polarizations. The effective refractive index for the fundamental guided mode  of the infrared (IR) pump light (labeled E$_{11}$), at a vacuum wavelength of 1064~nm, is shown as red curves. The effective refractive index of the next-higher transverse modes, E$_{ij}$, of the according second-harmonic at 532~nm wavelength is shown as green and blue curves, for quasi-$x$ (horizontally) and quasi-$y$ (vertically) polarized light (Figs.~\ref{fig:shg_phasematching}(a) and (b), respectively). It can be seen that modal phase matching involving the E$_{11}$ mode and a given polarization is expected only for a single, specific waveguide width where the IR dispersion curve crosses a specific SH dispersion (green) curve as  indicated with a black circle. According to the calculations, phase matching is expected with a waveguide width of about $w=0.65$~$\mu$m for the horizontally polarized transverse mode E$_{13}$, and about $w=0.68$~$\mu$m for the vertically polarized transverse mode E$_{21}$. 

\begin{figure}[htbp]
	\centering			
	\includegraphics[width=\linewidth]{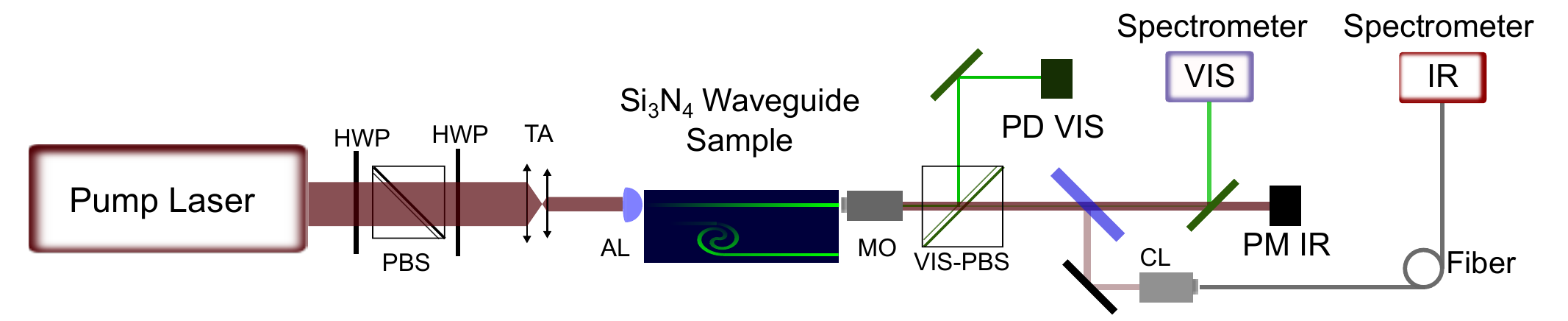}
	\caption{\label{fig:shg_setup}  Schematic of the experimental setup used for second-harmonic (SH) generation in Si$_3$N$_4$ waveguides. The infrared pump laser with a wavelength near 1~$\mu$m is sent through two half-wave plates (HWP), through a telescope arrangement (TA), and a polarizing beam splitter (PBS) and focused into a waveguide sample with an aspheric lens (AL). Output from the waveguide is collected with a microscope objective (MO), followed by a dichroic mirror (DM) that spectrally separates the fundamental IR from the second-harmonic. A photodiode (PD VIS) behind a polarizing beam splitter (VIS PBS) and a power meter (PM IR) record the SH power and IR radiation \textit{vs.} time. A small portion of the output (1\% of the SH and 16\% of the IR) is guided into two spectrometers to monitor the visible (VIS) and infrared (IR) spectra, using a collection lens (CL) and a large mode area fiber.}
\end{figure}

Figure \ref{fig:shg_setup} shows the experimental setup employing a mode-locked Yb-fiber laser operating at a wavelength of 1064~nm with a pulse duration of 6.2~ps and a repetition rate of 20~MHz (Toptica, PicoFYb 1064). The power of the laser and its polarization was controlled via two half-wave plates and a polarizing beam splitter. To maximize the input coupling, the round cross-section provided by the laser in a collimated beam was reduced from 4~mm to 2.5~mm with a telescope arrangement. Due to the high confinement of IR light in the waveguide, with a calculated effective mode area of around 0.5~$\mu$m$^2$, achieving maximally efficient input coupling would require a rather high numerical aperture ($\mathit{NA}\approx$1.25) whereas, due to availability, we used an aspheric lens with moderate numerical aperture ($\mathit{NA} = 0.68$, Thorlabs C330TMD-C). With this arrangement, the input coupling loss estimated from throughput measurements amounted to about 75\%. The output light from the other waveguide facet was collected via a microscope objective ($\mathit{NA}=0.74$, Mitutoyo x60) with an estimated output coupling loss of 43\% for the fundamental (as obtained from the measured collimated output from the microscope objective and power measurements with a large area power meter). 
The fundamental and SH powers were continuously monitored \textit{vs.} time. For monitoring the respective spectra, power fractions of about 16\% for the fundamental and 1\% for the second-harmonic power were sent to separate spectrometers. 

\begin{figure}[htbp]
	\centering	
	\begin{tabular}{@{}c@{}}
		\includegraphics[width=0.8\linewidth]{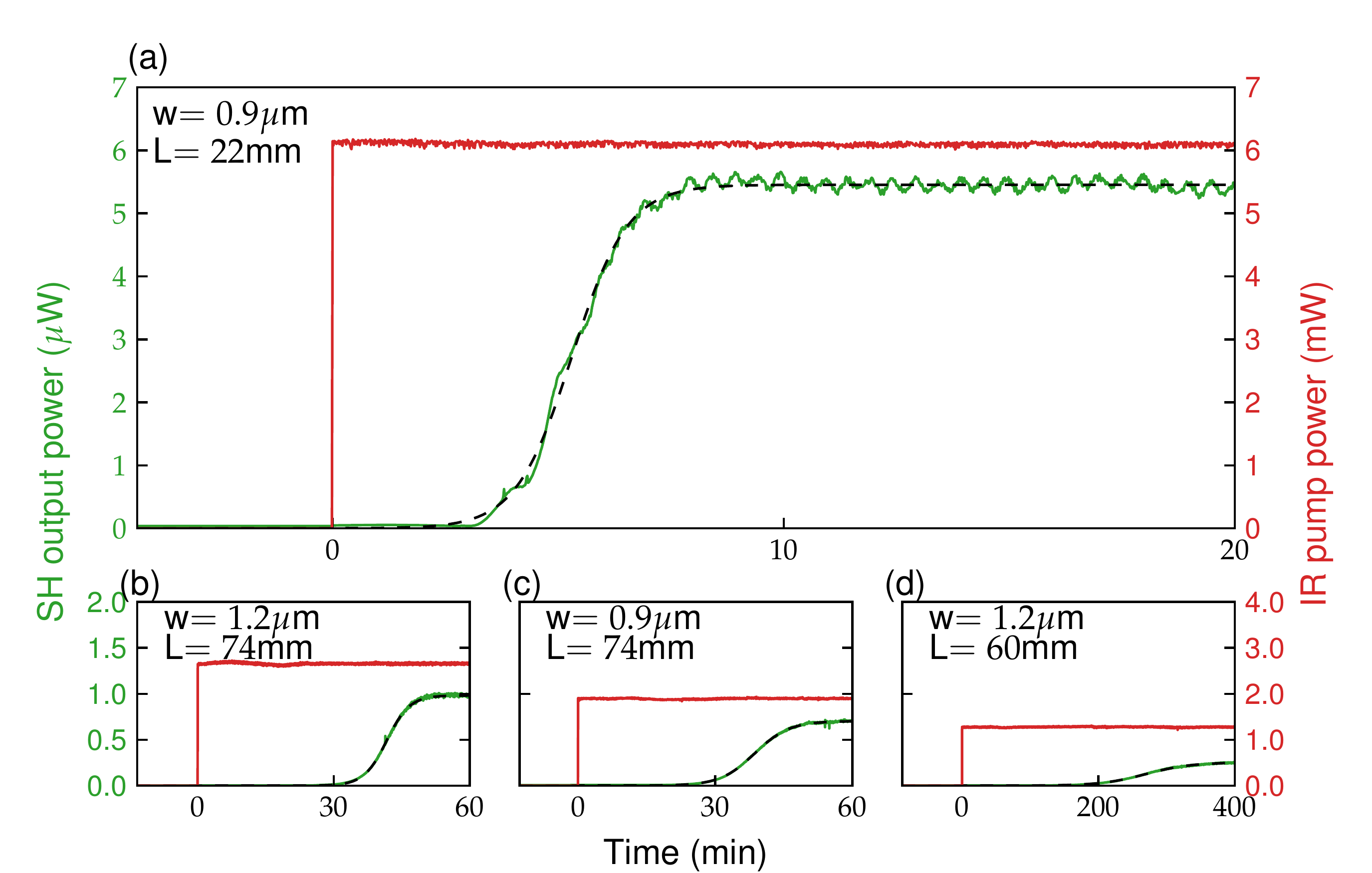}
	\end{tabular}
	\begin{tabular}{@{}c@{}}
		\includegraphics[width=0.8\linewidth]{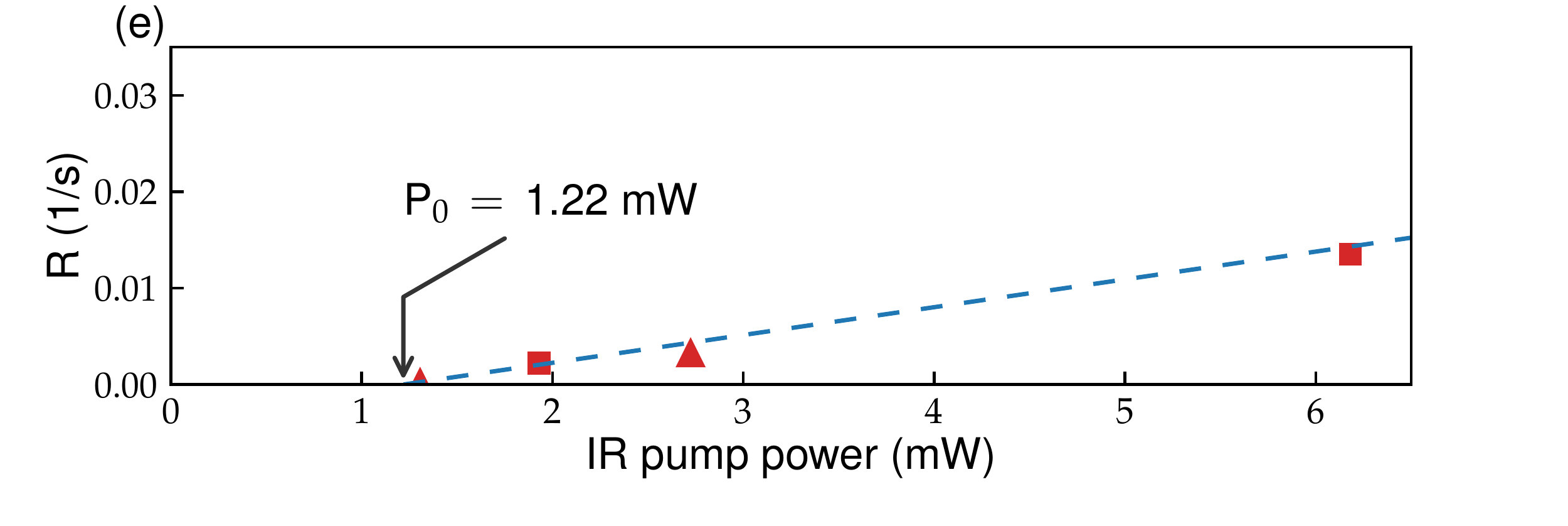}
	\end{tabular}
	\caption{\label{fig:shg_turnon2} 
		(\textbf{a}) to (\textbf{d}). Initial growth of the second-harmonic output power \textit{vs.} time (green traces) in four waveguides of different widths, $w$, and lengths, $L$. The red traces show the average infrared pump power in the waveguide. The dashed curves are least-square fits of the exponential function $f(t)$ (see main text) to the experimental data. (\textbf{e}). Shown is the rate of growth, $R$, where the SH output has reached a 1/e-fraction of its steady-state value, as retrieved from the fits to the data in (a-d), \textit{vs.} the waveguide-internal pump power. The triangular and square symbols represent growth rates as obtained with waveguide widths of  $w = 0.9$~$\mu$m and 1.2~$\mu$m, respectively. The dashed curve is a linear least-square fit to the data.} 
\end{figure}

In Figs.~\ref{fig:shg_turnon2}(a-d) several typical examples of time dependent measurements performed during the initialization are shown, \textit{i.e.}, when exposing the waveguides to pump radiation for the first time. 
The examples comprise two different waveguide widths ($w = 0.9$~$\mu$m and 1.2~$\mu$m), and three different waveguide lengths ($L= 74$, 60 and 22~mm). The power of the waveguide-internal pump laser, in these examples using horizontally polarized light, is shown as red traces and was kept constant after turn-on. The traces shown in green display the generated SH \textit{vs.} time. It can be seen that the SH output does not appear simultaneously with the pump laser turn-on. Instead, the SH builds up with a delay of minutes to hours, after which it rises within tens of seconds to minutes until reaching a steady state, with the sooner and steeper growth occurring at higher pump powers.

For comparison with the experimental data we applied a least-square fit using the exponential growth function as given in~\cite{balakirev_relaxation_1996},  $ f(t) = \frac{a}{1+(Rt_0-1)\,e^{-2tR}} $, where $a$, $t_0$ and $R$ are fit parameters.  In this expression, $a$ is the steady-state value, $t_0$ is the delay time and $R$ quantifies the growth rate of the SH output, \textit{i.e.}, the steepness of the slope at the moment that the output reaches a $1/e$-fraction of the steady-state value. The figure shows that, for all pump powers, the fit function matches the experimental data very well. For an evaluation of the power dependence, the values for the growth rate retrieved from the fits are displayed as data points in Fig.~\ref{fig:shg_turnon2}(e) \textit{vs.} the pump power, $P_p$. 
For a comparison, we used a least-square fit of a linear function of the pump power as $R= (P_p-P_0)/F$, where $F$ and $P_0$ are fit parameters. $P_0$ is the pump power where the growth rate becomes zero, indicating the existence of a minimum or threshold pump power required to obtain SHG. Figure~\ref{fig:shg_turnon2}(e) shows that the fit agrees well with the measured growth rates. The threshold pump power retrieved from the fit, $P_0$ = 1.22~mW, is indicated in Fig.~\ref{fig:shg_turnon2}(e) with an arrow.  

\begin{figure}[htbp]
	\centering
	\includegraphics[width=0.75\linewidth]{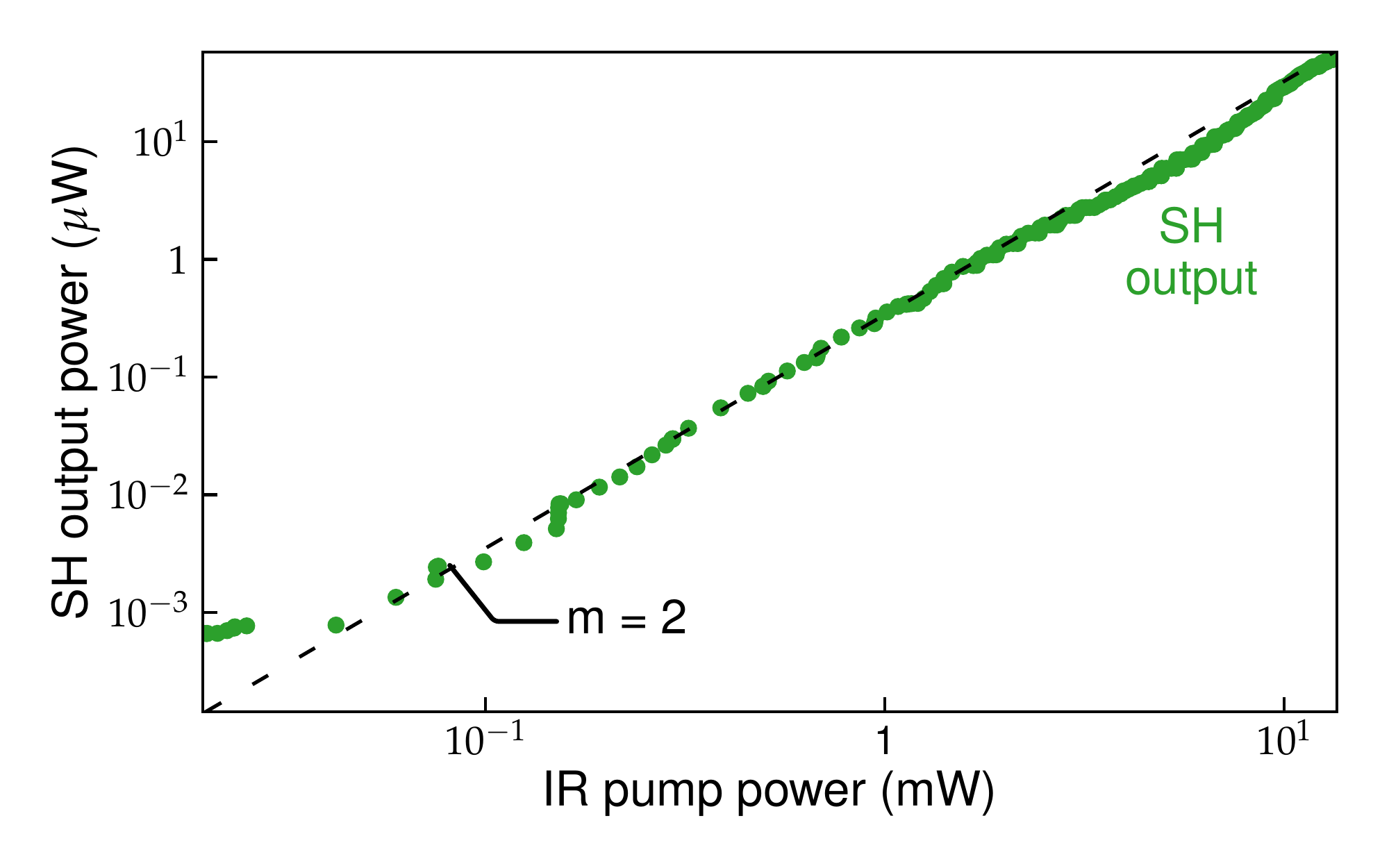}
	\caption{\label{fig:shg_09_22} Measured second-harmonic (SH) output as function of input infrared pump power. The used waveguide has a width of 0.9~$\mu$m, a height of 1.0~$\mu$m and a length of 22~mm. The dashed line is a quadratic fit curve (slope $m=2$).
	}
\end{figure}

In order to verify that the observed green output is indeed SHG and not radiation from a different process, such as fluorescence from impurity ions, the SH output power was measured \textit{vs.} the IR pump power. Expected is a quadratic growth with the pump power and that the output shows a well-defined polarization. The experiment was carried out by reducing the pump power towards zero, starting from the value that was used to generate a steady-state SH output, and also by increasing it from zero back to the maximum available power. In these measurements, performed after the initialization as described above, we observed that the SH power followed the pump power reduction or increase without any noticeable delay (sampling time 1~s). Also, the output was found linearly polarized, parallel to the polarization of the pump laser. The measured SH output \textit{vs.} pump power is displayed in Fig.~\ref{fig:shg_09_22}. A double-logarithmic plot is chosen for easy identification of power laws, where a quadratic dependence shows as a straight line with a slope, $m$, of 2. The dashed line shown in the figure is a least-square fit with a fixed slope  of $m = 2$ and with a variable offset as single fit parameter. It can be seen that the generated output power follows very well the expected quadratic dependency of second-harmonic generation. The agreement clearly proves that the investigated waveguides show a second-order nonlinear response.

\begin{figure}[btp]
	\centering
	\includegraphics[width=0.75\linewidth]{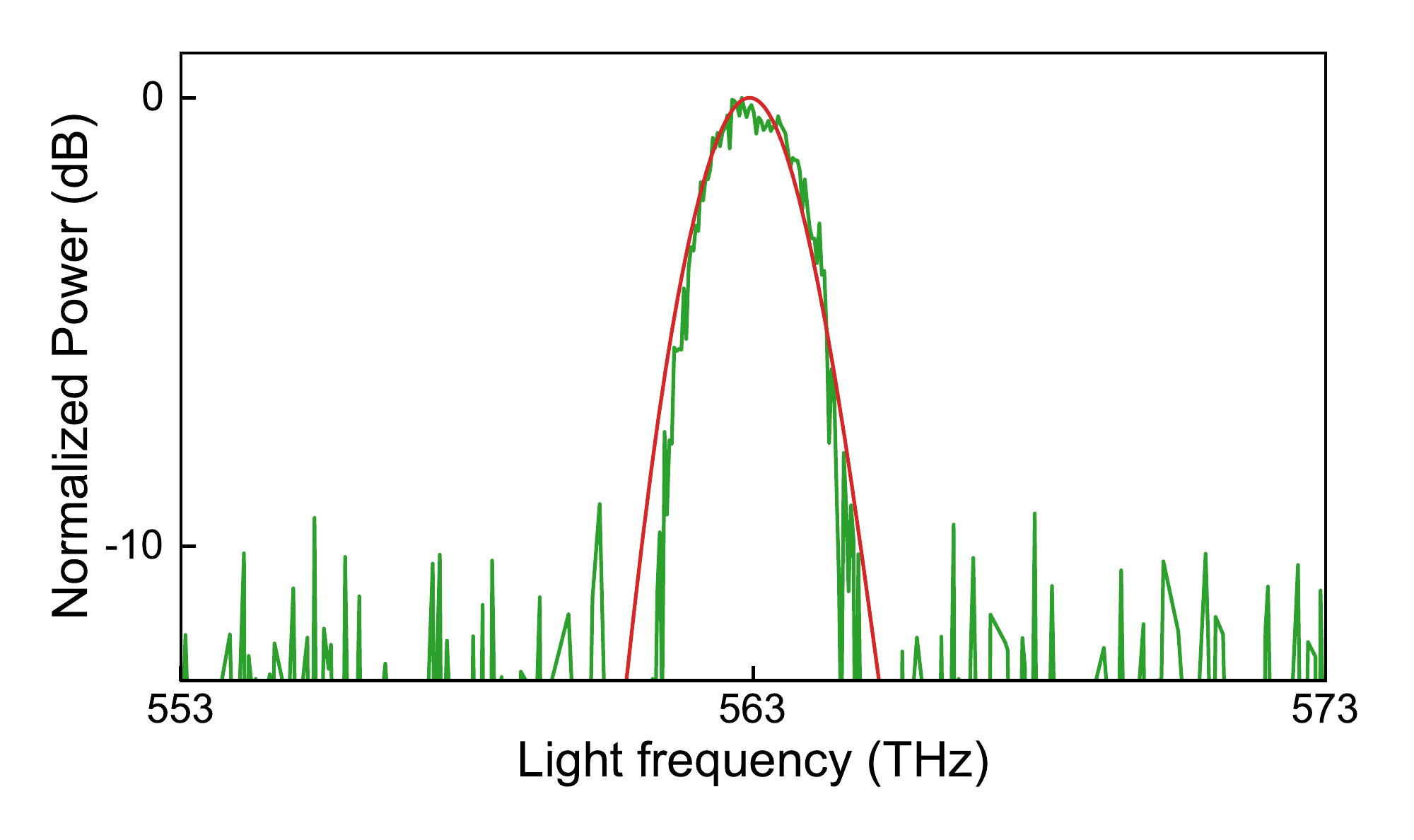}
	\caption{\label{fig:shg_spectra} Power spectra of the generated visible radiation (green trace) compared with the autocorrelated infrared pump power spectrum (red trace) in a logarithmic plot, with peak values normalized to 0~dB. The SH was generated in a waveguide with a $0.6\times1$~$\mu$m$^2$ cross-section and a length of 36 mm, with a waveguide-internal pump power of 4.8~mW.
	}
\end{figure}

In order to investigate whether the generated output involves components also of higher-order nonlinear response, specifically, self-phase modulation or four-wave mixing via the $\chi^{(3)}$-nonlinear susceptibility, we recorded the power spectra of the pump laser and the SH output for comparison with each other. If the output is solely dependent on the pump power with a square-law, one expects that the spectral shape of the SH output should match the convolution of the IR spectrum with itself (autocorrelation of the IR power spectrum). Figure \ref{fig:shg_spectra} compares an example of a measured and normalized power spectrum of the SH (green trace) with the normalized autocorrelation spectrum of the measured pump spectrum (red trace). It can be seen that the SH power spectrum fits the autocorrelated pump spectrum well, which excludes noticeable spectral influences of self-phase modulation and four-wave mixing.

\begin{figure}[htbp]
	\centering
	\begin{tabular}{ c c }                                                     
		\includegraphics[width=0.45\linewidth]{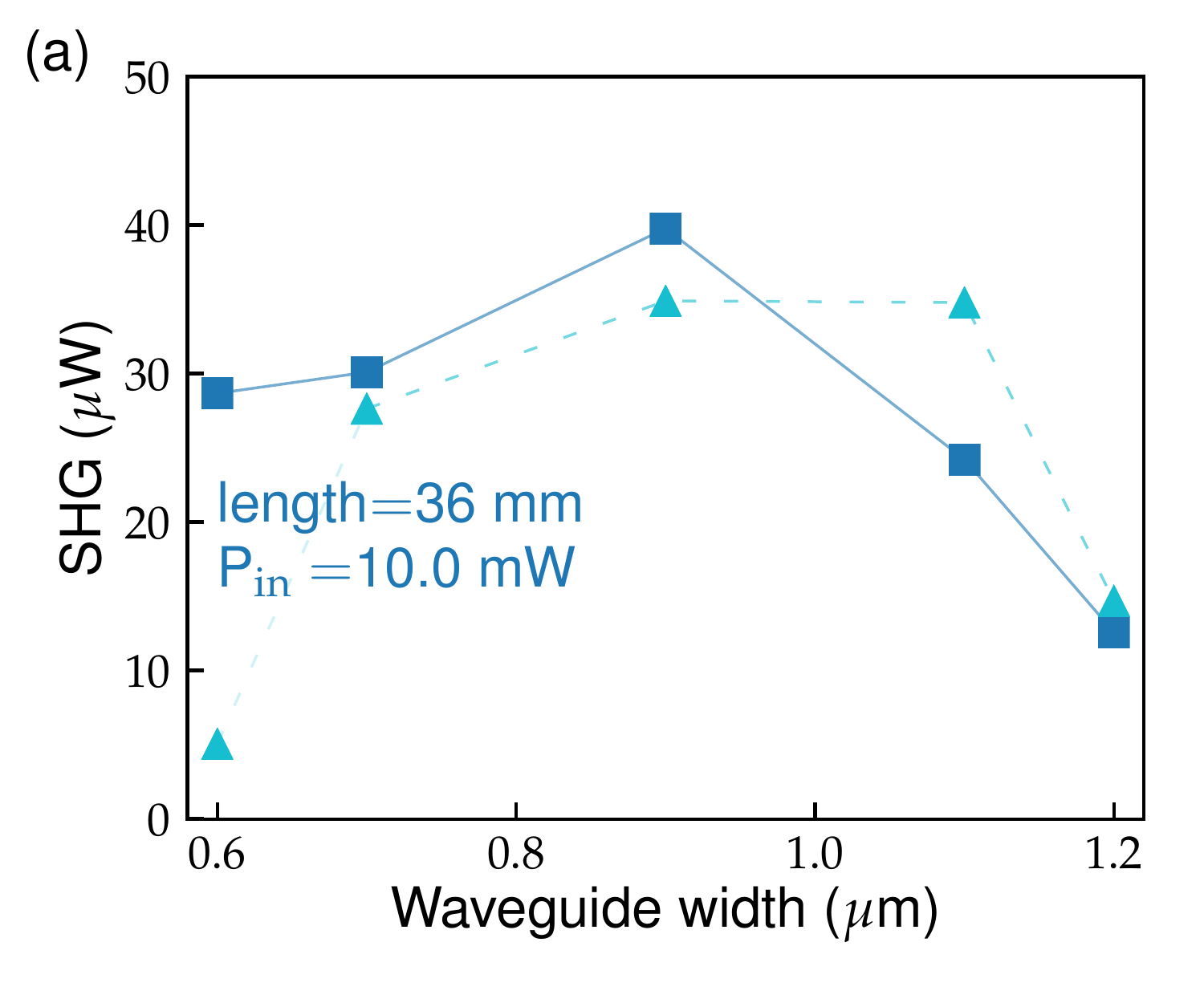}
		\includegraphics[width=0.45\linewidth]{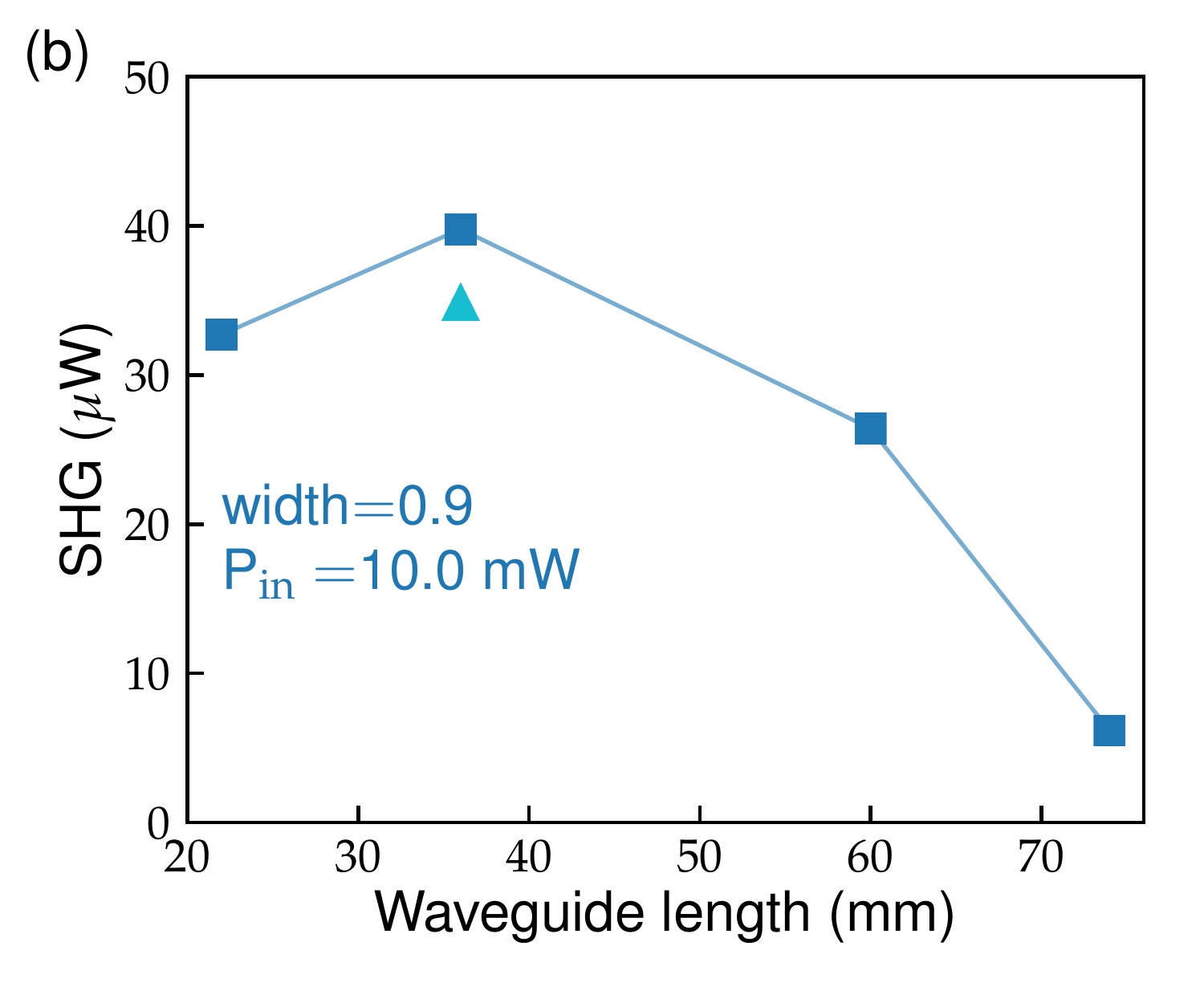}
	\end{tabular}
	\caption{\label{fig:shg_quad} \textbf{(a)} Measured second-harmonic (SH) output power for different waveguide widths with a fixed input power of 10~mW, horizontally polarized (squares) and vertically polarized light (triangles) for a waveguide length of 36~mm. \textbf{(b)} Measured second-harmonic power for different waveguide lengths with a fixed input power of 10~mW, for a waveguide with  $0.9\times 1$~$\mu$m$^2$ cross-section.} 
\end{figure}

A next measurement was performed to attempt clarifying what type of phase-matching mechanism was present that enables generation of the observed SH. Figure~\ref{fig:shg_phasematching} suggests that modal phase matching should be possible only with a single, specific value for the waveguide width. The experiments, however, showed that SH output is generated regardless of the waveguide width, such that equal phase velocities of transverse modes alone cannot explain the observed phase matching. For instance, SHG is observed with $w=0.9$~$\mu$m as well as with 1.2~$\mu$m (see Fig.~\ref{fig:shg_turnon2} and also Fig.~\ref{fig:shg_quad}(a)). To gain further information by identifying the spatial mode in which the SH is generated, we recorded the transverse intensity profile of the visible output in the far-field behind the waveguide, with a microscope objective and a CCD. To provide a large $\mathit{NA}$ for the recording, a microscope objective was placed ($\mathit{NA}=1.4$, 60x, \textit{Zeiss}) with immersion oil (index of 1.5) slightly closer than the focal distance, in order to maintain a diverging beam for a far-field measurement.

\begin{figure}
	\centering
	\includegraphics[width=1\linewidth]{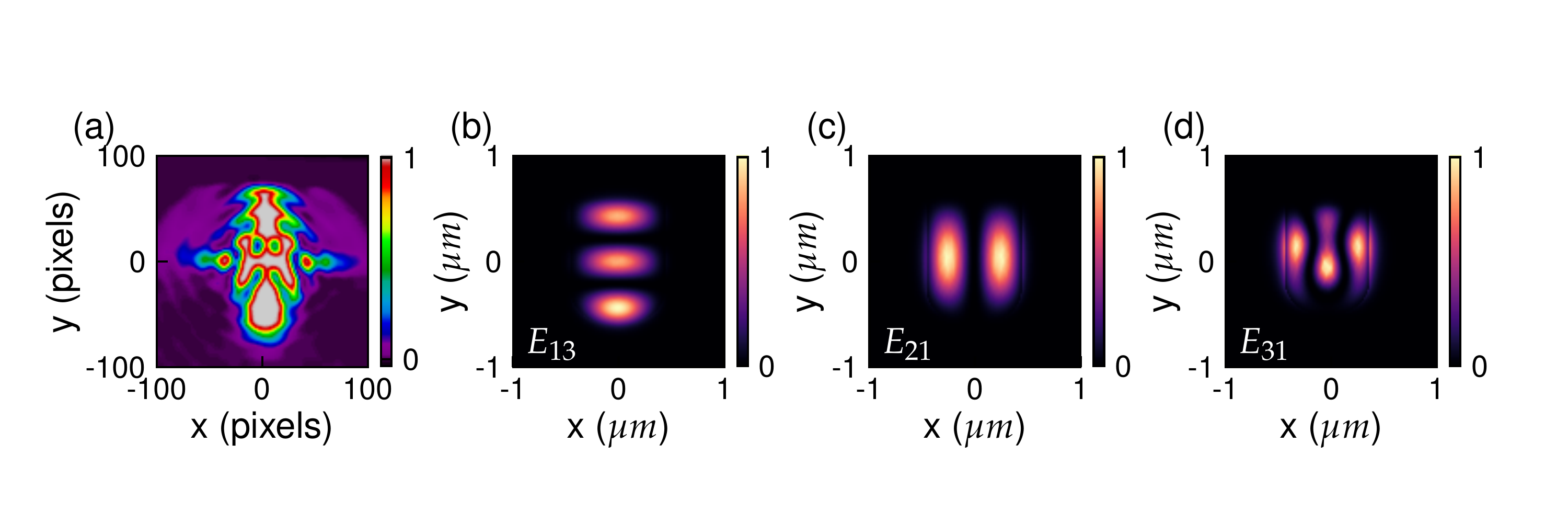}
	\caption{\label{fig:shg_mode} 
		Far-field intensity pattern formed by the second-harmonic output beam, normalized to the maximum intensity, recorded with a CCD behind an immersion objective \textbf{(a)} and calculated normalized intensity pattern for the waveguide modes E$_{13}$ \textbf{(b)}, E$_{21}$ \textbf{(c)} and E$_{31}$ \textbf{(d)}. The waveguide has a width of 0.7~$\mu$m and a height of 1~$\mu$m. The SH output and waveguide modes are horizontally polarized.
	}
\end{figure}

The CCD was positioned about 1~m beyond the microscope objective. The measured SH far-field intensity profile, normalized to the maximum intensity, for a waveguide with a cross-section of $0.7\times1$~$\mu$m$^2$ and length of 36~mm is shown in Fig.~\ref{fig:shg_mode}(a). Figure~\ref{fig:shg_mode} shows a transverse-moded structure with vertically three main lobes (the lower one and the higher one having the highest intensity, the central one distorted), and with horizontally two weaker side lobes. As the SH intensity profile could only be measured in the far-field, it is not possible to determine the modal decomposition of the SH to great detail. Further complications are some diffraction by the objective aperture and possible asymmetries introduced by the meniscus of the oil droplet present between waveguide facet and microscope objective\footnote{Asymmetries became clearly visible in the experiment when the objective was moved too far from the exit facet of the waveguide}. However, the main features in the measured pattern, such as number of lobes and symmetry, allows the identification of the main mode in which the SH is generated. 

To that end, Figs.~\ref{fig:shg_mode}(b-d) show the normalized intensity profiles of the waveguide modes E$_{13}$, E$_{21}$ and E$_{31}$, respectively, for the horizontal polarization. In calculating the transverse profiles of these eigenmodes, the actual waveguide geometry as determined from SEM measurements was used~\cite{epping_high_2015}. The waveguide core has rounded edges at the bottom, \textit{i.e.}, at the location of minimum $y$ value in Fig.~\ref{fig:shg_mode}, and this asymmetry is responsible for the higher intensity of the lower lobes in the profiles shown in Fig.~\ref{fig:shg_mode}. Qualitatively, the presence of a weak central spot with vertically stronger and horizontally weaker  side spots, the highest similarity was found with the E$_{13}$ mode, with some small contributions from either the E$_{21}$ mode, the E$_{31}$ mode or both. For reasons discussed below, we also expect that the SH radiation is dominantly generated in the E$_{13}$ mode for this waveguide.

Finally, in order to identify the optimum overall conditions for SHG we recorded the dependence of the SHG output \textit{vs.} the waveguide width, length, and polarization. Figure \ref{fig:shg_quad}(a) shows the measured power for five different waveguide core widths, 0.6, 0.7, 0.8, 0.9, 1.1, and 1.2~$\mu$m (height of 1~$\mu$m and length of 36~mm), using the same waveguide-internal pump power of 10~mW. Squares represent horizontally polarized light and triangles vertically polarized light. Figure \ref{fig:shg_quad}(b) shows the SH output power as a function of the waveguide length using a fixed waveguide core cross-section ($0.9\times1$~$\mu$m$^2$) and the same waveguide-internal pump power (10~mW).

A maximum SH output is found for a waveguide length of $L=36$~mm, and a cross-section of $0.9\times1$~$\mu$m$^2$ for horizontally polarized pumping. Regarding polarization, two effects were observed when changing the input polarization. Initially, \textit{i.e.}, immediately after injecting the other polarization, there was no SH output. Instead, a SH build-up took place with a certain delay and growth rate as described in Fig.~\ref{fig:shg_turnon2}. Thereafter, the SH output followed instantaneously any variation of the pump power. The second observation is that the polarization of the SH output was always parallel to that of the pump radiation. Independent of polarization effects it was observed that the second-order nonlinear response became erased when illuminating the waveguide with UV radiation (a PR-100, UVP Inc., with a 254~nm wavelength was used for ten minutes). After storing initialized waveguides in complete darkness at room temperature, typically for at least a week, we did not observe any degradation of the nonlinear response.

\section{Discussion}
The most basic expectation that the second-harmonic (SH) output power would follow a quadratic growth with the input pump power, and that the spectrum would have the same spectral bandwidth and shape as the autocorrelated pump spectrum, is confirmed by the experimental observations. However, contrary to what is expected, which is the absence of SH output, due to fabrication as amorphous materials, we observed second-harmonic generation, though this required an initialization process during which the SH is time dependent. This time dependency is a growth over time of the SH power towards a steady-state value, as long as the pump power is above a threshold value ($P_0$ as indicated in Fig.~\ref{fig:shg_turnon2}(e)). Unexpected is also that SHG was observed in all of the investigated waveguides, independent of their width, which excludes an explanation solely based on modal phase matching. Instead, during the initialization, an additional wave-vector, \textit{i.e.}, a matching spatial periodicity, appears to become available that compensates the remaining wave vector mismatch of modal phase matching. 

We address our observation of phase-matched SHG to the so-called coherent photogalvanic effect (CPGE), because of the excellent match of the fit functions to the measured temporal growth function and growth-rate \textit{vs.} power dependence in Figs.~\ref{fig:shg_turnon2}, \textbf{(a-e)}). The temporal fit function was derived by Balakirev \textit{et al.}~\cite{balakirev_relaxation_1996} for the CPGE in glass. The first observations of the CPGE in guided optics were made with phosphor-doped silica fibers~\cite{osterberg_dye_1986, stolen_self-organized_1987, mizrahi_test_1988, margulis_second-harmonic_1988, dianov_photoinduced_1995} and several phenomenological models were proposed~\cite{anderson_model_1991, sokolov_formation_1993, sokolov_phenomenological_1995, balakirev_relaxation_1996, chmela_first_1998}.

In brief, a first part of the CPGE is related to the third-order term in the expansion of Ohm's law, $\vec{j} = \sigma \vec{E}$, in powers of the electric field~\cite{sokolov_theory_1995}. Here, $\vec{j}$ is the current density, $\sigma$ is the conductivity and $\vec{E}$ a directed (DC) electric field generated by the light fields present in the waveguide. A second, underlying nonlinear effect is third-order optical rectification, which originates from the simultaneous presence of light at the fundamental frequency, $\omega$, and its second harmonic, at $2\omega$. The $\chi^{(3)}$-nonlinear rectification field induces a photocurrent density that is proportional to $\vec{E}(\omega)\vec{E}(\omega)\vec{E}^{*}(2\omega)$. This nonlinear DC current density possesses a spatial periodicity set by the wavenumber mismatch between the fundamental and second-harmonic fields ($\Delta k=2k(\omega)-k(2\omega)$). The current density leads to a spatial redistribution of electrons between long-lived intra-band trap sites~\cite{anderson_model_1991, dianov_photoinduced_1995, balakirev_anisotropy_2003}, which is the origin for generating an effective second-order nonlinearity. An equivalent picture is the creation of a nonlinear conduction via multi-photon excitation of electrons into the conduction band. Simultaneously, the optical rectification field created via the third-order susceptibility creates a charge relocation between the trap sites with a spatial pattern that provides self-organized quasi-phase matching. 

For a somewhat more detailed description, we make the simplification that all optical fields have the same linear polarization, in agreement with our experimental observation. The third-order nonlinear conductivity and susceptibility tensors are assumed to be isotropic and can be represented as scalars as the materials under consideration are amorphous. Accordingly, we treat all fields, current densities and tensors as scalar quantities.

We define the electric field of the fundamental input wave as 
\begin{equation}
E({\omega};\vec{r}) = A_{1}(z) E({\omega};x,y) cos( k(\omega) z + \varphi_{1}),
\end{equation}
and the second-harmonic field as 
\begin{equation}
E({2\omega};\vec{r}) = A_{2}(z)E({2\omega};x,y) cos( k(2\omega) z + \varphi_{2}).
\end{equation}
Here, $z$ is the propagation coordinate, $k(\omega)$ and $k(2\omega)$ are the wavenumber of fundamental and SH, and $\varphi_{1}$ and $\varphi_{2}$ are phase offsets for the fundamental and SH fields. $A_{1}(z)$ and $A_{2}(z)$ are the amplitudes of the field distributions along the propagation direction for the fundamental and the SH waves, $E({\omega};x,y)$ and $E({2\omega};x,y)$ represent the transverse field distributions, respectively. 

Electrons, promoted to the conduction band via multi-photon absorption, are driven away from their original locations by a static (DC) electric field, $E_{DC}$, that is generated via the third-order nonlinearity (third-order optical rectification),
\begin{equation} \label{eq_rectification}
P^{(3)}_{DC}(0;\vec{r}) = \varepsilon_0 \chi^{(3)}(0=\omega+\omega-2\omega)E({\omega};\vec{r})E({\omega};\vec{r})E^{*}({2\omega};\vec{r}).
\end{equation}
Equation~\ref{eq_rectification} shows that the optical rectification field exhibits a spatial modulation set be the wavevector mismatch between the fundamental and second harmonic fields that is contained in the $z$-dependence of $E({\omega};\vec{r})$ and $E({2\omega};\vec{r})$. Consequently, the associated nonlinear photocurrent density $j_{ph}$ given by~\cite{dianov_photoinduced_1995, balakirev_kinetics_2003} 
\begin{equation} \label{eq_photocurrent}
j_{ph}(\vec{r})= C(x,y) |A_{1}(z)|^2 A_{2}(z) |E({\omega};x,y)|^2 E^{*}({2\omega};x,y) \cos(\Delta k z + \Delta\varphi)
\end{equation}
exhibits the same spatial modulation. In Eq.~\ref{eq_photocurrent}, $C(x,y)$ is the effective photogalvanic coefficient~\cite{sokolov_phenomenological_1995, sokolov_theory_1995, balakirev_kinetics_2003}, $\Delta k = 2k(\omega)-k(2\omega)$ is the wavenumber mismatch, and $\Delta\varphi$ is a constant. The direction of the current follows the transverse polarization of the inducing light fields. Via this photocurrent and due to the presence of trap-sites, a long-lived charge grating is written into the material which remains present when the optical fields are turned off. 

In the presence of light, the charge grating continues to develop, but the development ceases when the transversely orientated space-charge field, $\mathcal{E}_{DC}$, associated with the charge grating balances the optical rectification field, 
\begin{equation} \label{eq_opposingForce}
\varepsilon_0 \chi^{(1)} \mathcal{E}_{DC} = -\varepsilon_0 \chi^{(1)} E_{DC} = -P^{(3)}_{DC}(0;\vec{r}).
\end{equation}
Alternatively expressed, the development of the charge grating will cease when the space-charge field has increased to a level where 
\begin{equation}
\mathcal{E}_{DC}(\vec{r})= -j_{ph}(\vec{r})/\sigma_{\mathrm{eff}},
\end{equation} 
where  $\sigma_{\mathrm{eff}}$ is the effective conductivity of the material in the presence of the optical fields. This conductivity is composed of the third-order nonlinear conductivity $\sigma^{(3)}$ and of a background conductivity, also named dark conductivity, which has a low value for the case of amorphous glass~\cite{sokolov_phenomenological_1995}. The spatial period of $\mathcal{E}_{DC}(\vec{r})$ along $z$ is again given as Eq.~\ref{eq_photocurrent} by the fundamental \textit{vs.} SH wavenumber mismatch.

Finally, the fundamental light field, $E(\omega,\vec{r})$, together with the static field, $\mathcal{E}_{DC}$, generates a second-harmonic polarization via the material's third-order nonlinear response, of which the transverse component is given by
\begin{equation}
\mathcal{P}^{(3)}({2\omega;z}) =
\varepsilon_0\chi^{(3)}(z)\mathcal{E}_{DC}(\vec{r}) E(\omega;\vec{r}) E(\omega;\vec{r}).
\end{equation}
This third-order response at the second-harmonic frequency can be seen as an effective second-order polarization based on an effective second-order nonlinearity, as given by~\cite{dianov_photoinduced_1995} 
\begin{equation} \label{eq_chi_eff}
\chi^{(2)}_{\mathrm{eff}}(\vec{r}) = \chi^{(3)}\mathcal{E}_{DC}(\vec{r}).
\end{equation}
In Eq.~\ref{eq_chi_eff}, the induced $\mathcal{E}_{DC}$, and thus also $\chi^{(2)}_{\mathrm{eff}}$ assumes the form of a spatial grating along $z$ with the periodicity set by $\Delta k$.  The spatial structure of $\chi^{(2)}_{\mathrm{eff}}$ is then similar to a periodically poled second-order nonlinear crystal that provides quasi-phase matching~\cite{fejer_quasi-phase-matched_1992}, in this case for second-harmonic generation.

\subsubsection*{Phenomenological description of second harmonic mode selection}
As described above, the CPGE provides quasi-phase matching independent of the wavenumber mismatch. Therefore, a second-harmonic output should be generated independent of the chosen waveguide width which was indeed experimentally observed. However, for the same reason, the generation should then also be possible independent of the specific type of transverse mode in which the second-harmonic field is generated. For instance, SHG should also be generated in the fundamental mode, E$_{11}(2\omega)$ by the fundamental IR mode, E$_{11}(\omega)$. In contrast to this, we have observed the SH output to be generated predominantly in a specific higher-order (transverse) mode shown in Fig.~\ref{fig:shg_mode}(a), \textit{i.e.}, in the E$_{13}$ mode (Fig.~\ref{fig:shg_mode}(b)) with only some small E$_{21}$ (Fig.~\ref{fig:shg_mode}(c)) or E$_{31}$ (Fig.~\ref{fig:shg_mode}(d)) modes.
This rises the question, why the dominant SH mode is E$_{13}$. In order to discuss possible reasons, in the following we present a phenomenological description of details in the generation of the spatial charge distribution expected in our case.

To generate an initial photocurrent via an optical rectification field, an initial SH seed wave is required. Initially, the seed wave might posses only a small amplitude. Nevertheless, the nonlinear feedback, enabled by the photocurrent-generated charge grating responsible for the effective second-order nonlinearity that leads again to second-harmonic generation, will eventually lead to an exponential growth starting from the seed wave. An initial SH seed wave can be present, \textit{e.g.}, from surface SHG near the interface between core and cladding~\cite{boyd_nonlinear_2013}, near local material anisotropies~\cite{anderson_model_1991}, or by photoionization of defects, \textit{i.e.}, of electrons from dangling bonds~\cite{warren_si_1993, sokolov_theory_1995}. 

Two mechanisms can be responsible for the nonlinear optical generation of a conductivity, required for the coherent photogalvanic effect. The first possibility is excitation of electrons via three-photon absorption from the valance band to the conduction band~\cite{balakirev_anisotropy_2003}. This possibility is, however, not likely in our experiments since the bandgap of stoichiometric Si$_3$N$_4$ (4.9~eV) is even larger than the energy of four pump photons (4.6~eV). The second mechanism is via dangling bonds in stoichiometric silicon nitride. In LPCVD Si$_3$N$_4$ two main types of dangling bonds are possible, silicon (K-centers, having an energy of $\sim$2.6~eV above the upper edge of the valence band) and nitrogen (N-centers, having an energy of $\sim$0.3~eV above the upper edge of the valence band)~\cite{krick_electrically_1988}. These dangling bonds allow photoexictation of electrons into the conduction band. As the dangling bond defects are most stable in their charged state~\cite{robertson_nature_1995} they also act as trap sites and provide the mechanism to create long-lived charge gratings in the core of the waveguide. 

The establishment of the charge grating may be described and understood as follows. The time dependency of the SH output is dominated by the excitation of charges by the pulsed laser (ps timescale), while the recombination time of charges to a vacancy is rather long ($\sim$0.1~s)~\cite{dianov_transient_1994}. Due to the high repetition rate of the laser pulses present over a significant illumination time, the excited electrons can be pushed away by the optical rectification field from volume elements where the product of the light fields, $E^{*}(2\omega;\vec{r}) \cdot E(\omega;\vec{r})^2 $, is large, towards trap sites in the direction given by the rectification field. If the optical rectification field is not strong enough, recombination will be dominant and no charge grating is written, \textit{i.e.}, the average IR laser power should be above a threshold, which is in agreement with our experimental observation (\textit{cf.} Fig.~\ref{fig:shg_turnon2}(e)).  

\begin{figure}[tb]
	\centering
	\includegraphics[width=1\linewidth]{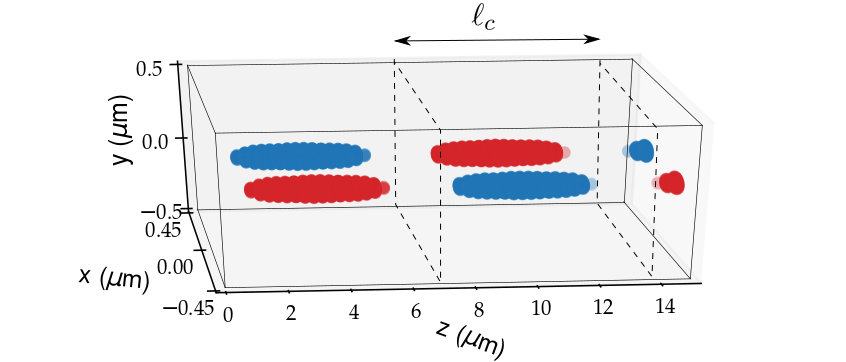}
	\caption{ \label{fig:shg_3Dschematic} 
		Normalized charge distribution in the waveguide core induced by the coherent photogalvanic effect. The generating optical fields are assumed to be the fundamental mode E$_{11}(\omega;\vec{r})$ for the IR field and the E$_{13}(2\omega;\vec{r})$ mode for the SH field, both polarized along the $x$-direction. For clarity only charge values with $|\rho|$ above 80\% of the maximum value are plotted, where red and blue represents positive and negative charge, respectively. The sign of the effective $\chi^{(2)}$ is then given from blue to red. The horizontal arrow indicates the coherence length for SHG, $\ell_c = \frac{\pi}{\Delta k}$, in absence of quasi-phase matching.
	}
\end{figure}

To illustrate the shape of charge distribution written in the core of the waveguide, we have calculated the product $E(\omega;\vec{r})^2 E^{*}(2\omega;\vec{r}) \propto \mathcal{E}$ and used Poisson's equation, simplified to $\rho(x) \propto d\mathcal{E}/dx$, where $\rho$ is the charge density, to determine from the transverse field the according charge density. Figure \ref{fig:shg_3Dschematic} shows the example of a charge distribution as generated by the pump radiation in the E$_{11}$ mode and by SH in the E$_{13}$ mode, both having a horizontal polarization. For better clarity of presentation, only charge levels with $|\rho|$ in the range of 80\% - 100\% of the maximum charge are shown. Positive and negative charge are indicated by red and blue, respectively. Figure~\ref{fig:shg_3Dschematic} shows that the overlapping pump and SH fields generate a charge grating where the longitudinal periodicity defines a grating period, $\Lambda =2\ell_c = \frac{2\pi}{\Delta k}$, where $\ell_c$ is the coherence length for second-harmonic generation in absence of quasi-phase matching. This grating period is what is required for quasi-phase matched (QPM) generation of the second harmonic field.

\begin{figure}
	\centering
	\includegraphics[width=0.75\linewidth]{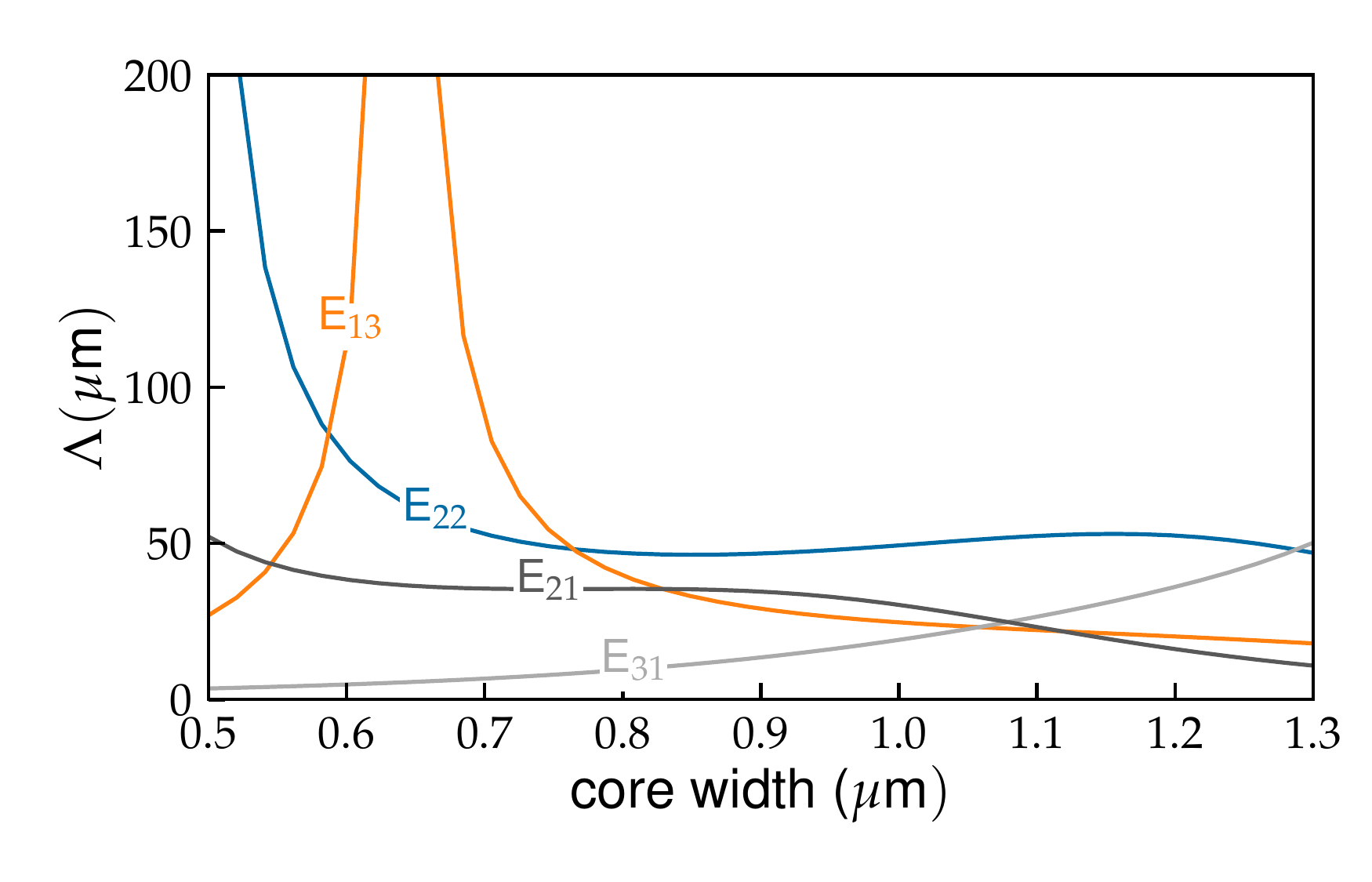}
	\caption{\label{fig:shg_period} QPM period required for SHG in the various different modes shown in Fig.~\ref{fig:shg_phasematching} with the infrared pump power in the E$_{11}$ mode. All modes are horizontally polarized. As can be seen, E$_{13}$ has the largest period for the range of waveguide widths discussed in Fig.~\ref{fig:shg_phasematching}.
	}
\end{figure}

At this point it can be discussed, why not all of the allowed spatial modes are equally likely to generate a SH output. We suggest that the answer lies in the very different charge patterns that correspond to the different pump and SH mode combinations. As can be seen in Fig.~\ref{fig:shg_3Dschematic} the longitudinal grating period, $\Lambda$, is set by the wavenumber mismatch between the mode of the IR pump, here taken to be the fundamental E$_{11}$ mode, and the mode of the SH field. A smaller phase mismatch, $\Delta k$, and consequently a larger grating period, $\Lambda$, results in a stronger growth with distance for the SH in the waveguide~\cite{boyd_nonlinear_2013}. Initially, the SH field might be emitted in various modes. But then, during the initiation phase of exponential growth, the various resulting charge gratings will compete with each other to establish the charge grating in the waveguide. The mode with the largest growth, \textit{i.e.}, with the smallest phase mismatch, will become the dominant mode due to the nonlinear feedback process described above.  
Figure \ref{fig:shg_period} depicts the QPM period, $\Lambda$, as a function of the width of the waveguide core for various modes with horizontal polarization. 

Figure~\ref{fig:shg_period} shows that, for a waveguide width of $w=0.7$~$\mu$m \textit{i.e.}, for the waveguide used to record the SH beam profile in Fig.~\ref{fig:shg_mode}(a), the required QPM period, $\Lambda$, is rather short for most of the SH modes, and consequently these modes will only experience a moderate growth. Only certain modes, here E$_{13}$, can be quasi-phase matched with a long QPM period and will experience a far larger growth, \textit{i.e}, will become the dominant mode. This is in agreement with the measured SH beam profile (see Fig.~\ref{fig:shg_mode}(a).

As a related conclusion, from Fig.~\ref{fig:shg_period}, it appears more likely that the observed SH output is a E$_{13}$-E$_{21}$ superposition (with dominating E$_{13}$) than having a E$_{31}$ contribution. The E$_{31}$ contribution, that may be present in the observed SH intensity distribution, requires an even shorter QPM period at $w=0.7$~$\mu$m and is expected to have a smaller growth. Based on these phase-matching arguments,  Fig.~\ref{fig:shg_period} would imply that SH generated in the E$_{22}$ mode grows faster than that generated in the E$_{21}$ mode. However, this mode is not expected to grow due to the zero on-axis intensity of the E$_{22}$ mode, which creates a poor modal overlap of this mode with the E$_{11}$ mode of the IR pump.    

\begin{figure}[bt]
	\centering
	\includegraphics[width=0.75\linewidth]{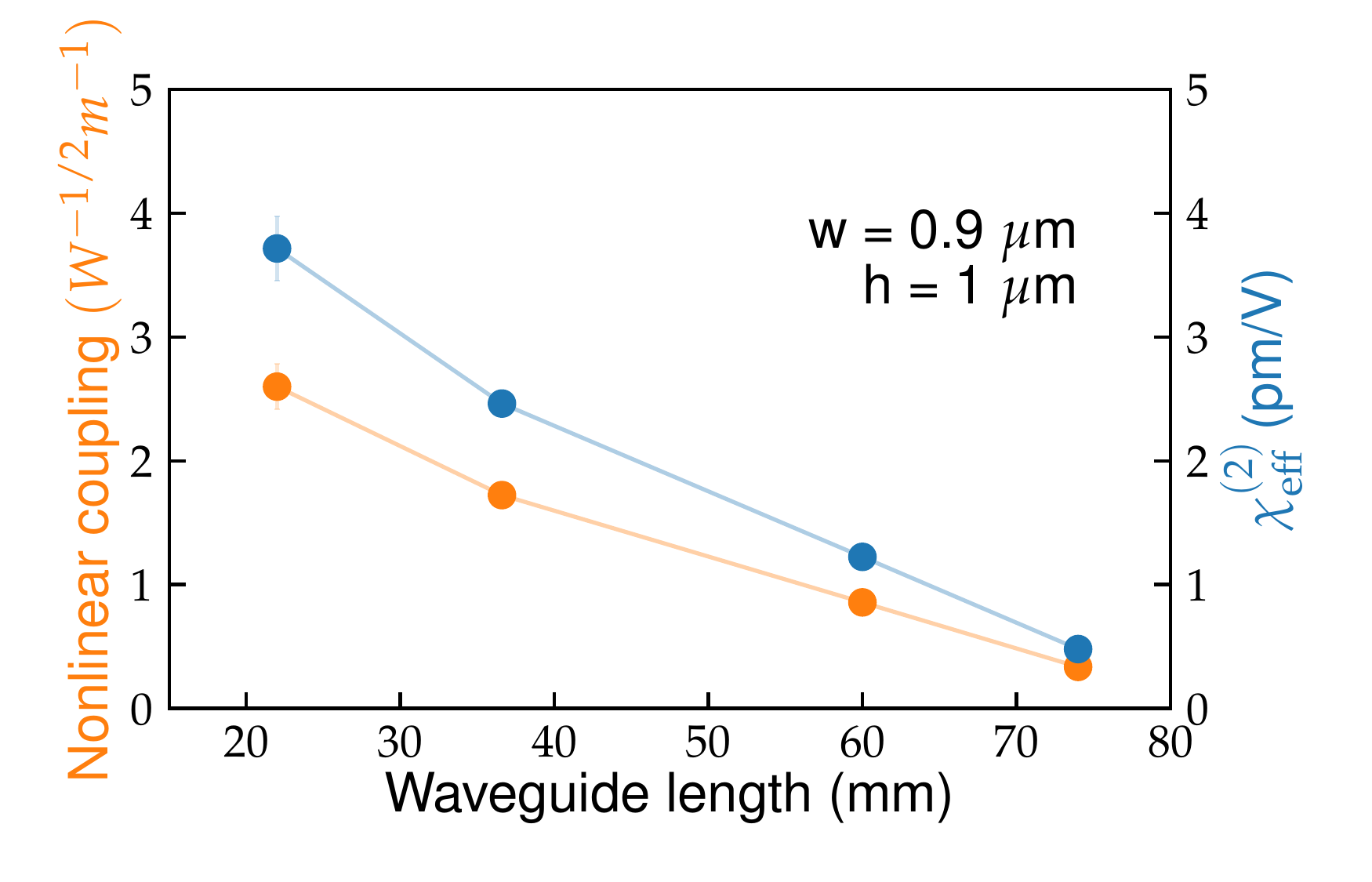}
	\caption{\label{fig:shg_kappachi} Nonlinear coupling (orange markers) and effective $\chi^{(2)}$ (blue markers) as a function of the length of the waveguide. The core of the waveguide has a cross-section of $0.9\times1$~$\mu$m$^2$. The shown error bar (0.2~pm/V half-width) seen at the highest data points represents the statistical uncertainty of the fit in Fig.~\ref{fig:shg_09_22} and the experimental uncertainty in measuring the output coupling efficiency at the waveguide exit facet.}
\end{figure}

In order to quantify the effective nonlinear susceptibility than can be provided by the CPGE in Si$_3$N$_4$ waveguides, we evaluate the maximum power conversion efficiency, $\eta$, which was obtained for a width of $w=0.9$~$\mu$m (core cross-section $0.9\times1$~$\mu$m$^2$). The conversion efficiency is defined as
\begin{equation}\label{shg_eq_eta}
\eta = P(2\omega)/P_0(\omega) = \kappa^2 P_0({\omega}) L^2,
\end{equation}
where $P({2\omega})$ and $P_0({\omega})$ are the generated SH power and the input IR power, respectively, $\kappa$ is the coefficient for nonlinear coupling between the fundamental and SH waves, and $L$ is the length of the waveguide. Equation \ref{shg_eq_eta} can be used to determine $\kappa$ from measured powers and the waveguide length. 
Assuming perfect quasi-phase matching and no pump depletion, the obtained value for $\kappa$ can be used to determine the effective second-order nonlinear susceptibility~\cite{suhara_waveguide_2003}, $\chi_{\mathrm{eff}}^{(2)}$ as
\begin{equation} \label{shg_eq_chieff}
\chi^{(2)}_{\mathrm{eff}} \, = \, 2\sqrt{ 
	\left(\frac{\kappa}{ \varepsilon_0}\right)^2 \frac{2(n(\omega)_{\mathrm{eff}})^2 \, n({2\omega})_{\mathrm{eff}}}{(2\omega)^{2}} \left(\frac{\mu_0}{\varepsilon_0}\right)^{-3/2} {S_{\mathrm{eff}}}  },
\end{equation}
where $n_{\mathrm{eff}}({\omega})$ and $n_{\mathrm{eff}}({2\omega})$ are the refractive indices for the IR and SH waves, while $S_{\mathrm{eff}}$ is the effective cross-section of the modal overlap of the IR and the SH fields~\cite{suhara_waveguide_2003}. In our case, the horizontally polarized pump radiation is injected in the E$_{11}$ mode and the SH is dominantly generated in the E$_{13}$ mode with the same polarization. Figure \ref{fig:shg_kappachi} shows the experimentally obtained nonlinear coupling, $\kappa$, as a function of the length of the waveguide. Also shown is the corresponding effective second-order susceptibility as calculated using Eq. \ref{shg_eq_chieff}, assuming a pure E$_{13}$ mode for the SH output. Figure~\ref{fig:shg_kappachi} shows that the maximum effective value for $\chi^{(2)}$ obtained in this way is about 3.7$\pm$0.2~pm/V. We note that taking into account a small SH contribution in the E-21 mode would lead to a larger effective susceptibility. 

\section{Summary and conclusions}
For the first time, we have observed second-harmonic generation in LPCVD-grown stoichiometric Si$_{3}$N$_4$ waveguides. Second-harmonic output was generated independent of the width of the waveguide core cross-section and required a build-up, a delay in the order of several minutes up to hours and an exponential growth to a steady state value, depending on the IR pump power and waveguide core cross-section. These findings are consistent with the coherent photogalvanic effect being responsible for the building up of a second-order nonlinearity and quasi-phase matching. With the available pump laser pulses, the investigated Si$_3$N$_4$ waveguides provide a maximum conversion efficiency of 0.4\% for an input power of 13~mW (peak power 105~W, pulse energy of 0.65~nJ). This is the highest second-order conversion efficiency achieved to our knowledge in an integrated SiN platform. The measured conversion efficiency corresponds to a high effective second-order susceptibility, $\chi^{(2)}_{\mathrm{eff}}$ = 3.7$\pm$0.2~pm/V, which is smaller than that from highly Si-rich SiN thin films ($d_{\mathrm{eff}}=5.9$~pm/V, corresponding to a $\chi^{(2)}_{\mathrm{eff}}=11.8$~pm/V)~\cite{kitao_investigation_2014} as concluded from X-ray photoelectron spectroscopy (XPS), and slightly larger than that achieved with periodic gratings in thin SiN films ($\chi^{(2)}$ = 2.5~pm/V)~\cite{ning_efficient_2012} or in SiN ring resonators ($\chi^{(2)}$ < 0.04~pm/V)~\cite{levy_harmonic_2011}, both calculated from frequency conversion measurements. 

Numerous applications are already based on $\chi^{(1)}$-gratings in waveguides such as reconfigurable Bragg filters~\cite{marpaung_integrated_2013}, optical switching~\cite{stegeman__1996, suhara_waveguide_2003}, mode conversion~\cite{chen_second_2013} or optical storage~\cite{calafiore_holographic_2014}. The presence of an effective $\chi^{(2)}$-grating in amorphous $\text{Si}_{3}\text{N}_4$ waveguides may lead to the development of further functionalities such as parametric down-conversion~\cite{cap_self-organized_2003}, all optical signal processing~\cite{stegeman__1996}, and possibly also self-referencing of frequency combs that exploits the simultaneous presence of $\chi^{(2)}$ and $\chi^{(3)}$ in low-loss Si$_3$N$_4$ waveguides~\cite{telle_carrier-envelope_1999, jones_carrier-envelope_2000, mayer_frequency_2015}.

\section*{Funding}
This research is supported by NanoNextNL (6B-Functional Nanophotonics), a micro- and nanotechnology consortium of the Government of the Netherlands and 130 partners, and the Netherlands Organization for Scientific Research, NWO, (STW project 11358), which is partly funded by the Ministry of Economic Affairs.

\section*{Acknowledgments}
We thank Albert van Rees and Richard Mateman from LioniX International BV. for designing the lithography mask and for cleanroom processing and fabrication of the samples, respectively.


\end{document}